\documentclass[12pt]{iopart}

\usepackage{graphicx}
\usepackage{bm}

\begin{document}
\title[]{The oscillating two-cluster chimera state in non-locally coupled phase oscillators}
\author{Yun Zhu$^1$, Yuting Li$^1$, Mei Zhang$^2$, Junzhong Yang$^1$}
\address{1 School of Science, Beijing University of Posts and
Telecommunications, Beijing, 100876, People's Republic of China}
\address{2 Physics Department, Beijing Normal University, Beijing, 100875, People's Republic of China}
\ead{zhuyun@bupt.edu.cn}

\begin{abstract}
We investigate an array of identical phase oscillators non-locally
coupled without time delay, and find that chimera state with two
coherent clusters exists which is only reported in delay-coupled
systems previously. Moreover, we find that the chimera state is
not stationary for any finite number of oscillators. The existence
of the two-cluster chimera state and its time-dependent behaviors
for finite number of oscillators are confirmed by the theoretical
analysis based on the self-consistency treatment and the
Ott-Antonsen ansatz.
\end{abstract}


\maketitle

\section{Introduction}
An array of identical oscillators has been used to model a wide
range of systems, such as neural networks, convecting fluids,
laser arrays and coupled biochemical oscillators. These systems
exhibit rich collective behaviors including synchrony and
spatiotemporal chaos \cite{1,2,3,4}. Most of the earlier
theoretical works on these systems assume either local coupling
(nearest-neighbor interactions) or global coupling (infinite-range
interactions); a third type named non-locally coupling began to be
explored in the past years, which is somewhere between local
coupling and global coupling. In non-local coupled systems, the
oscillators interact with all others and the strength between
oscillators varies with the distance between them.

Chimera state is a spatiotemporal pattern in which some of the
identical oscillators are coherent and synchronous while others
remain incoherent \cite{5,6,7,9,10}. They usually appear in
systems with non-local coupling and could only be built for proper
initial conditions \cite{15}. Chimera state does not relate to the
partially synchronized states observed in populations of
nonidentical oscillators with dispersive frequencies in which the
splitting of the population roots in the inhomogeneity of the
oscillator themselves and the intrinsically fastest or slowest
oscillators remain desynchronized. Its emergence cannot be
ascribed to a supercritical instability of the spatially uniform
oscillation, because it occurs even if the uniform state is
stable. Chimera may be a paradigm to study unihemispheric sleep in
neurology which states a fact that many creatures sleep with only
half of their brain while the other half is still active at the
same time. \cite{mat06}.

In the year 2002, Chimera state was first reported by Kuramoto and
his colleagues \cite{5,6} when simulating the non-locally coupled
complex Ginzburg-Landau equation. They showed that identical
oscillators with non-locally symmetrical coupling could
self-organize into chimera states. Soon, spiral wave chimera
\cite{13,14} was discovered in two-dimensional arrays of
non-locally coupled oscillators. In succession, Abrams and
Strogatz \cite{7} found an exact solution for this state in a ring
of phase oscillators coupled by a cosine kernel. Recently, two
interesting findings on chimera state are reported. Firstly, in
the study of the non-locally coupled oscillators with time delay
\cite{11,12}, clustered chimera state that has spatially
distributed phase coherence separated by incoherence with adjacent
coherent regions in antiphase, was found. The observed clustered
chimera state in these systems is stationary and its pattern does
not change with time. Secondly, Abrams and Strogatz \cite{8} found
a breathing chimera state in a model consisting of two interacting
subpopulations of oscillators. Pikovsky and Rosenblum \cite{16}
considered oscillators ensembles consisting of several
subpopulations of identical units, with a general heterogeneous
coupling between subpopulations, through which they acquired
quasiperiodic chimera states. Laing \cite{17,19} summarized
chimera states in several heterogeneous networks of coupled phase
oscillators, in the mean time, he analyzed chimera state applying
the Ott-Antonsen ansatz \cite{18} in one-dimensional and
two-dimensional systems. He pointed out that, in one-dimensional
system, when parameter heterogeneity is introduced, a breathing
chimera state exists.

In this work, we study a one-dimensional array of non-locally
coupled identical phase oscillators. We find that, in the absence
of time delay, a two-cluster chimera state could exist. We also
find that, in the absence of parameter heterogeneity, the
two-cluster chimera state is not stationary but oscillating.
Different from Laing's results \cite{18}, we find that the
oscillation of the two-cluster chimera state only exists for the
system with a finite number of oscillators. Both the two-clustered
chimera state and finite size oscillations of the chimera state in
this model are analyzed based on the Ott-Antonsen ansatz.

\section{Model}

The array of non-locally coupled phase oscillators can be
described in a concise form as
\begin{eqnarray}\label{eq:1}
    \frac{\partial\phi}{\partial
    t}=\omega-\int_{-k}^kG(x-x^{\prime})\sin[\phi(x,t)-\phi(x^\prime,t)+\alpha]dx^\prime.
\end{eqnarray}
Here, $\phi(x,t)$ is the phase of the oscillator at position $x$
at time $t$. The space variable $x$ is in the range $[-k,k]$
$(0<k\leq\pi)$. The periodic boundary condition is imposed for
$k=\pi$, otherwise the no-flux boundary condition is imposed on
the system. $\omega$ is the natural frequency (same for all
oscillators), which plays no role in the dynamics. Without losing
generality, we can set $\omega=0$. The angle $\alpha$
$(0\leq\alpha\leq\frac{\pi}{2})$ is a tunable parameter. The
kernel $G(x-x^\prime)$ provides non-local coupling between
oscillators. $G(x)$ is non-negative, even, decreasing with $|x|$
along the array, and normalized to have unit integral. Following
Abrams and Strogatz \cite{7}, we make use of the cosine kernel
\begin{eqnarray}\label{eq:kernel}
 G(x)=\frac{1}{2(k+A\sin k)}(1+A\cos x)
\end{eqnarray}
where $0\leq A\leq 1$. When $k=\pi$, Abrams and Strogatz found a
chimera state with only one coherent cluster \cite{7}. However, we
find a novel chimera state which has two coherent clusters and is
oscillating for any finite number of oscillators. As mentioned
above, clustered chimera state is only found in the time-delay
coupled systems \cite{11,12} and oscillating chimera state is
found in the systems with subpopulations \cite{8,16} or the
systems with parameter heterogeneity [i.e., $\alpha=\alpha(x)$]
\cite{17}.

Let $\Omega$ denotes the angular frequency of a rotating frame
whose dynamics are simplified as much as possible, and let
$\theta=\phi-\Omega t$ denotes the phase of an oscillator relative
to this frame. The key idea behind the analysis of chimera stats
is the introduction of a mean-field-like quantity, namely, a
complex order parameter $Re^{i\Theta}$ \cite{5,6,7} which is
defined as
\begin{eqnarray}\label{eq:3}
    R(x,t)e^{i\Theta(x,t)}=\int_{-k}^{k}G(x-x^\prime)e^{i\theta(x^\prime,t)}dx^\prime
\end{eqnarray}
Then Eq (1) becomes
\begin{eqnarray}\label{eq:4}
    \frac{\partial\theta}{\partial
    t}=\omega-\Omega-R\sin(\theta-\Theta+\alpha).
\end{eqnarray}

For stationary state, $R$ and  $\Theta$ are time-independent and
only depend on space variable $x$. Let $\Delta=\omega-\Omega$
where $\Omega$ is the angular velocity for the oscillators in
coherent regions, the oscillators with $\Delta\leq R$ are in
coherent regions and are phase-locked to
$\theta=\arcsin(\frac{\Delta}{R})+\Theta-\alpha$ \cite{5,6,7}.

\section{Simulate and results}

\begin{figure}
\includegraphics[width=6.0in]{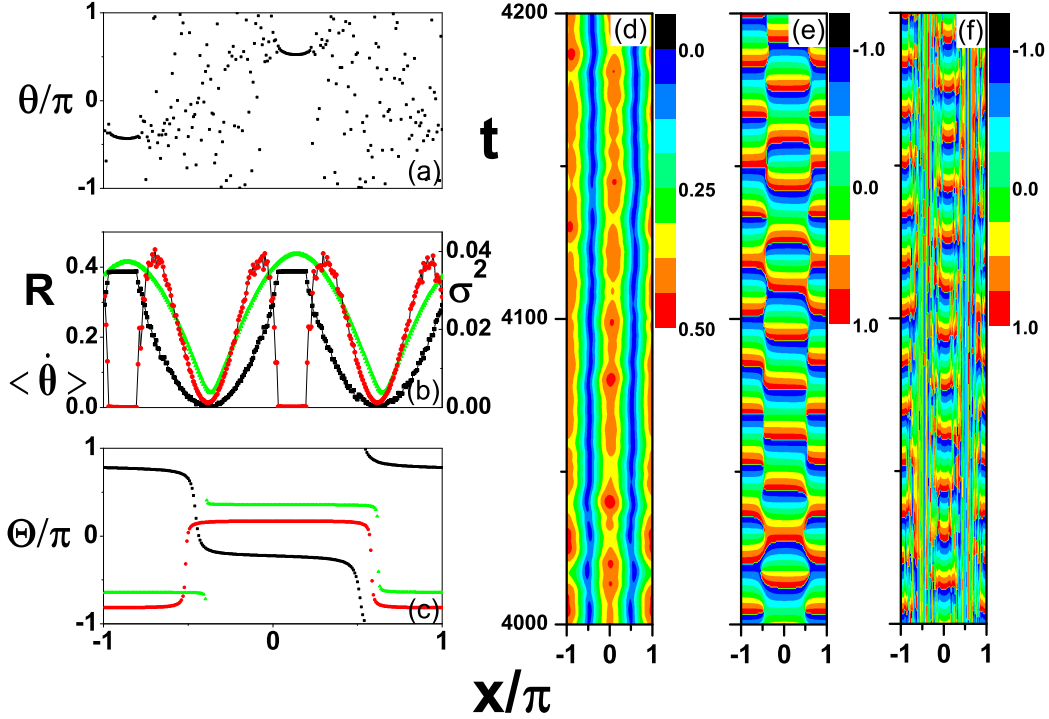}
\caption{\label{pi}(color online) (a) Phase pattern for
two-cluster chimera state when the steady state is reached. Eq.
(1) is integrated using the Runge-Kutta method with fixed time
step $dt=0.1$ with oscillator number $N=256$, $k=\pi$,
$\beta=0.10$ and $A=0.995$. (b) The triangle (green) symbol is the
modulus $R$ of the complex order parameter, the square (black)
symbol is the distribution of $\langle \dot{\theta}(x)\rangle$ of
individual oscillators averaged over 200 time units. The circle
(red) symbol is the fluctuation $\sigma^2(x)$ of $\dot\theta(x)$.
(c) three types of distribution of the phase $\Theta$ of the order
parameter at different time. (d), (e) and (f) show the contour
graphs of $R(x)$, $\Theta(x)/\pi$ and $\theta(x)/\pi$,
respectively. The horizontal axis is position $x$ and vertical
axis is time $t$. }
\end{figure}

For parameters  $\beta=\frac{\pi}{2}-\alpha=0.10$, $A=0.995$,
$k=\pi$, the system with $N=256$ phase oscillators could evolve to
a two-cluster chimera state under the initial conditions as
follows:
\begin{eqnarray}\label{intial condition}
\phi(x,0)=\left\{
\begin{array}{l l}
2\pi re^{-2.76x^2}, & x\leq 0\\
2\pi re^{-2.76x^2}+\pi, & x>0
\end{array}\right.
\end{eqnarray}
where $r$ is a random variable from a uniform distribution of
$[-0.5,0.5]$. As shown in Fig. 1(a), there exist two coherent
clusters in which all oscillators are synchronized. Oscillators in
the same cluster are nearly in phase yet in different clusters are
in antiphase. On the other hand, the oscillators between the two
coherent clusters are de-synchronized and their phases are
randomly distributed in $[-\pi,\pi]$.
\begin{figure}
\includegraphics[width=6.0in]{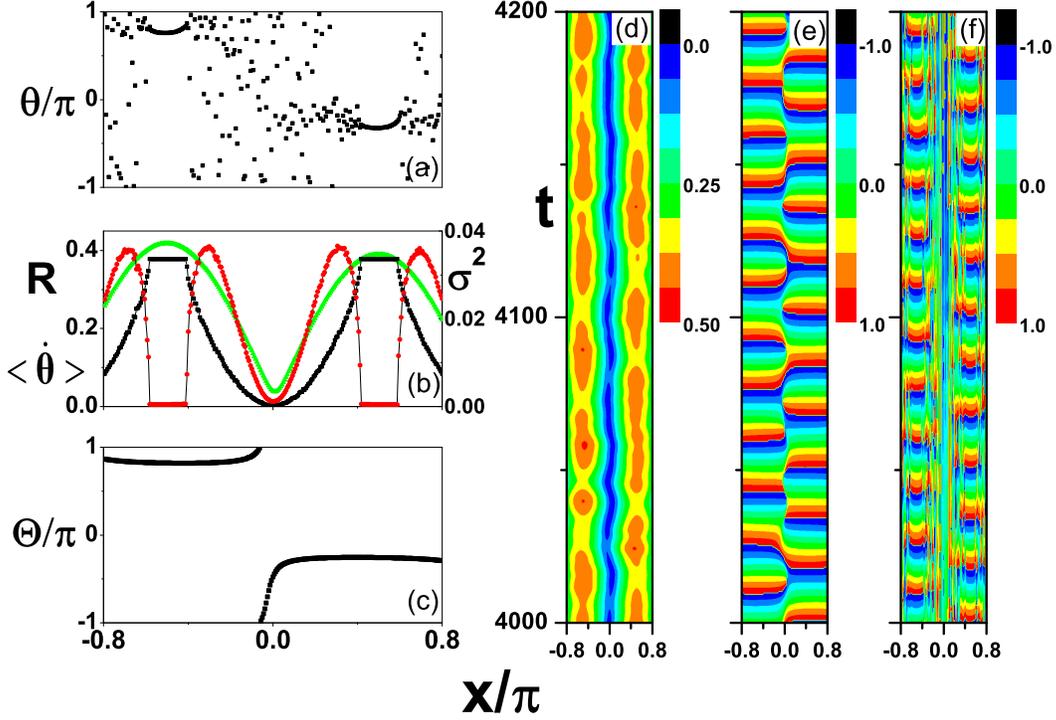}
\caption{\label{fig2}(color online) (a) Phase pattern for
two-cluster chimera. The parameters $N=256$, $k=0.8\pi$,
$\beta=0.10$ and $A=0.995$. (b) The triangle (green) symbol is the
modulus $R$ of the complex order parameter, the square (black)
symbol is the distribution of $\langle \dot{\theta}(x)\rangle$ of
individual oscillators averaged over 200 time units. The circle
(red) symbol is the fluctuation $\sigma^2$ of $\dot\theta(x)$. (c)
the distribution of the phase $\Theta$ of order parameter. (d),
(e) and (f) show the contour graphs of $R(x)$, $\Theta(x)/\pi$ and
$\theta(x)/\pi$, respectively.}
\end{figure}
Then we consider three quantities characterizing a chimera state:
the modulus $R$ of the complex order parameter at an arbitrary
time, the angular velocity $\langle \dot{\theta}(x)\rangle$
averaged in a time interval of 200 units, and the fluctuation of
the instantaneous angular velocity which is defined as
$\sigma(x)=\sqrt{\langle (\dot{\theta}(x)-\langle
\dot{\theta}(x)\rangle)^2\rangle}$. The quantities against the
locations of oscillators are presented in Fig. 1(b). Clearly,
there are two plateaus on the curve of $\langle
\dot{\theta}(x)\rangle$ which refer to the coherent clusters in
the chimera state. The zero $\sigma$ in the coherent clusters
means that the oscillators in the coherent clusters all move on
the same instantaneous angular velocity. Further, nonzero $\sigma$
outside the coherent clusters refers to the fluctuation of angular
velocities for the oscillators outside the coherent clusters and
indicates desynchronization. Fig. 1(b) reveals two features on $R$
for the two-cluster chimera state. Firstly, the oscillators can be
divided into two domains which join at the minimum of $R$ and the
curve of $R$ against the locations of oscillators does not show
symmetry about the minimum of $R$. Further explorations show that
$R$ is a function of time. As shown in Fig. 1(d) where the
spatiotemporal evolution of $R$ is featured, $R$ is oscillating in
each coherent cluster. Especially, when $R$ reaches its minimum in
one domain, $R$ in the other one reaches its maximum. Secondly,
the boundaries of the coherent clusters are not determined by the
condition of $\Delta=R(x)$ and the coherent regimes are narrower
than those expected according to $\Delta=R(x)$ in most of the
time, which are different from the stationary chimera state. The
reason for this observation roots in the time-dependent order
parameter $R$. Actually, for a forced phase oscillator obeying
$\dot{\theta}=\omega+R(t)\sin\theta$, the synchronization of the
phase oscillator by the force just requires the phase of the
oscillator to be confined within $(0,2\pi)$ but not to a fixed
value, which leads the onset of the synchronization of the
oscillator not to obey the condition of $\Delta=R(t)$ and the
onset of synchronization strongly depends on the details of the
functional form of $R(t)$. To be noted, even though $R$ is
time-dependent, the oscillators in the coherent clusters still
have the same angular velocity which does not fluctuate as
exhibited by zero $\sigma$. The spatiotemporal evolution of $R$ in
Fig. 1(d) shows another feature: the two-cluster chimera state
displays an irregular motion along the ring, for example, the
locations of coherent clusters vary with time. Similar phenomenon
is also observed for the chimera state with a single cluster
\cite{omel2010}. Furthermore, the snapshot and the time evolution
of $\Theta$ presented in Fig. 1(c) and (e) show that $\Theta$ is
almost uniform in each domain except for those near the junction
between domains and there is a phase difference of $\pi$ for
$\Theta$ in different domains. To be stressed, the features on
$\Theta$ revealed by Fig. 1(c) and (e) could be used as a more
general scheme for the initial conditions to generate an
oscillating two-cluster chimera state. That is, the initial
conditions in Eq. \ref{intial condition} could be changed to be
$\phi(x,0)=\pi$ for $x\leq 0$ and $\phi(x,0)=0$ for $x>0$. To get
a better illustration, we present the evolution of $\theta(x)$ in
Fig. 1(f) which shows that, resulting from the oscillation of $R$,
the territories of the coherent clusters alter with time.
Furthermore, it should be pointed out that, though a two-cluster
chimera state could be generated for the system characterized by
Eq. \ref{eq:1} in the absence of time delay under proper initial
conditions, we have not detected other chimera states with more
than two coherent clusters.
\begin{figure}
\includegraphics[width=5.0in]{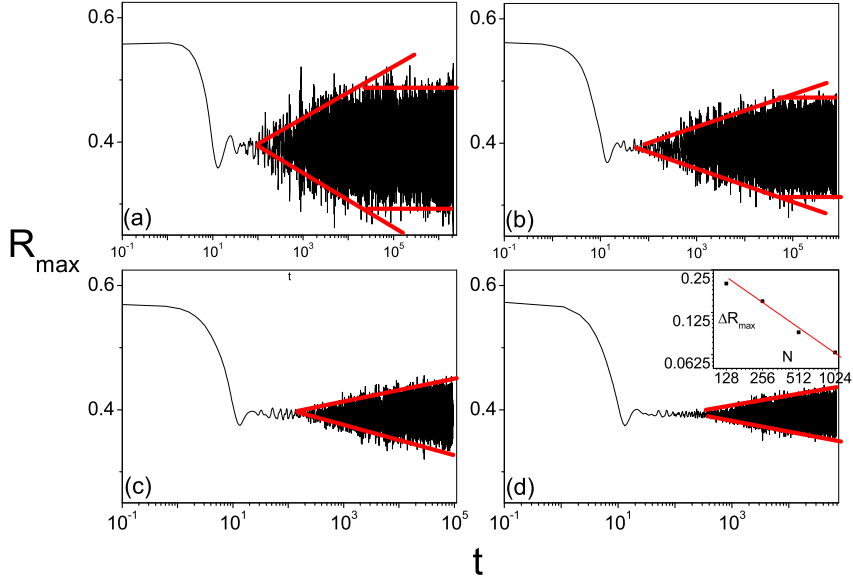}
\caption{\label{Rmax} (color online) The maxima of $R$ in one
domain varies with time for different number of oscillators. (a,
b, c, d) are for $N=128$, $256$, $512$ and $1024$, respectively.
Other parameters: $A=0.995$, $\beta=0.1$ and $k=0.8\pi$. The (red)
lines denote the growth of $R$ in the transient and the upper or
lower bound of the maxima of $R$ in the steady state. The inset in
(d) shows the amplitude of the oscillation of $R_{max}$, $\Delta
R_{max}$, against the system size $N$, which indicates a power law
of $\Delta R_{max}\sim N^{-0.5}$.}
\end{figure}

The two-cluster oscillating chimera state can exist when
$k\neq\pi$. In comparison with the case with $k=\pi$ where
oscillators locate on a ring, here the oscillators locate on a
chain. We take $k=0.8\pi$ as an example. The results are given in
Fig. \ref{fig2}. The differences with those in Fig. 1 are that the
pattern of two-cluster chimera state becomes stationary in space
and, no matter where the chimera state is initialized, it will
adjust its pattern to be symmetrical about the center of the chain
where the minimum of $R$ appears.

The analysis above are made on the system with $N=256$. One
question is how the observed two-cluster oscillating chimera state
depends on the number of oscillators. For this aim, we focus on
one domain and record the maximum value of $R$ in this domain at
any time instance (we denote it as $R_{max}$). The time-dependent
behavior of the two-cluster chimera state can be reflected by the
time evolution of $R_{max}$. For example, a constant $R_{max}$
indicates a stationary chimera state, otherwise an oscillating
one. To avoid the influences induced by the irregular motion of
the chimera state along the ring in the system with periodic
boundary condition, we let $k=0.8\pi$. The results for different
$N$ are presented in Fig. \ref{Rmax}. One remarkable feature
revealed by the figure is that, before the oscillating chimera
state is established, the system first evolves to a two-cluster
chimera state which looks like a stationary one due to the weak
oscillation of $R_{max}$. However, the "stationary" state is not
stable and its instability leads to the appearance of an
oscillating chimera state. Interestingly, the stationary value of
$R_{max}$ is independent of the size of the system and the time
consumed for the system to build an oscillating state becomes much
longer as the number of the oscillator $N$ increases. Another
feature in Fig.\ref{Rmax} is that larger $N$ seems to weaken the
oscillation of $R_{max}$ in the two-cluster oscillating  chimera
state, which is prominent by the comparison between Fig.
\ref{Rmax} (a) ($N=128$) and (b) ($N=256$). Since the transient
time to build an oscillating two-cluster chimera state for large
$N$ becomes extremely long, we just give a rough estimate based on
the data presented in the inset of Fig. \ref{Rmax}(d) that the
oscillation amplitude $\Delta R_{max}\sim N^{-0.5}$, which means
that the two-cluster oscillating chimera state becomes a
stationary one in the thermodynamic limit, which is different from
the Laing's results \cite{18}.

\section{Analysis}

\begin{figure}
\includegraphics[width=4in]{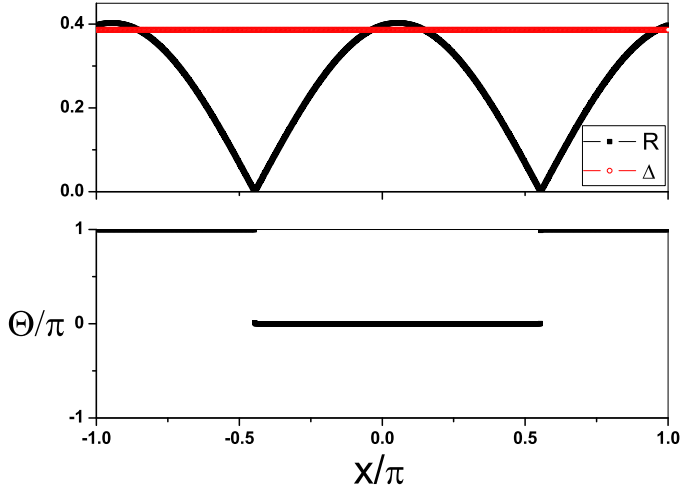}
\caption{\label{tukb}(color online) The modulus $R$ and the phase
$\Theta$ of the order parameter and $\Delta$ by solving of Eq.
(\ref{KBFInal}) via an iterative scheme as presented in the
context. (a) The square (black) symbol denotes the distribution of
$R$ and the circle (red) the value of $\Delta$. (b) Distribution
of $\Theta$. Other parameters: $A=0.995$, $\beta=0.1$ and $k=\pi$.
}
\end{figure}

The above results can be understood theoretically. First, the
two-cluster stationary chimera state in the thermodynamic limit
can be explained in terms of Kuramoto-Battogtokh self-consistency
equation\cite{5} as follows:
\begin{eqnarray}\label{KB}
R(x)e^{i\Theta(x)}&=&e^{i\beta}\int_{-\pi}^{\pi}G(x-x^\prime)e^{i\Theta(x^\prime)}
\times\frac{\Delta-\sqrt{\Delta^2-R(x^\prime)^2}}{R(x^\prime)}dx^\prime
\end{eqnarray}
note that there are three unknown quantities (the real-valued
functions $R(x)$, $\Theta(x)$ and the real number $\Delta$) in
terms of the assumed choices of $\beta$ and the kernel
$G(x-x^\prime)$. We take $k=\pi$ as an example. Considering the
patterns in Figs. 1(b) and (c), the modulus $R$ and the phase
$\Theta$ of the order parameter approximately satisfy
\begin{eqnarray}\label{symmetry}
\left\{
\begin{array}{l l}
R(x+\pi)=R(x)\\\Theta(x+\pi)=\Theta(x)+\pi.
\end{array}\right .
\end{eqnarray}
We substitute Eqs. (\ref{symmetry}) and (\ref{eq:kernel}) into Eq.
(\ref{KB}) and get
\begin{eqnarray*}
R(x)e^{i\Theta(x)}=e^{i\beta}\int_{-\pi}^0\frac{1}{2\pi}[1+A\cos(x-
x^\prime)e^{i\Theta(x^\prime)}H(x^\prime)dx^\prime
\end{eqnarray*}
\begin{eqnarray}\label{KB2}
+e^{i\beta}\int_{0}^{\pi}\frac{1}{2\pi}[1+A\cos(x-
x^\prime)]e^{i\Theta(x^\prime)}H(x^\prime)dx^\prime,
\end{eqnarray}
where $H(x)=(\triangle-\sqrt{\Delta^2-R(x)^2})/R(x)$. Under the
transformation  $x^\prime-\pi\rightarrow x^\prime$ in the second
term of the right hand side of Eq. (\ref{KB2}), the
self-consistency equation changes into
\begin{eqnarray}\label{KBFInal}
R(x)e^{i\Theta(x)}=e^{i\beta}\int_{-\pi}^0\frac{A\cos(x-x^\prime)}{\pi}
e^{i\Theta(x^\prime)}H(x^\prime)dx^\prime.
\end{eqnarray}
\begin{figure}
\includegraphics[width=5.0in]{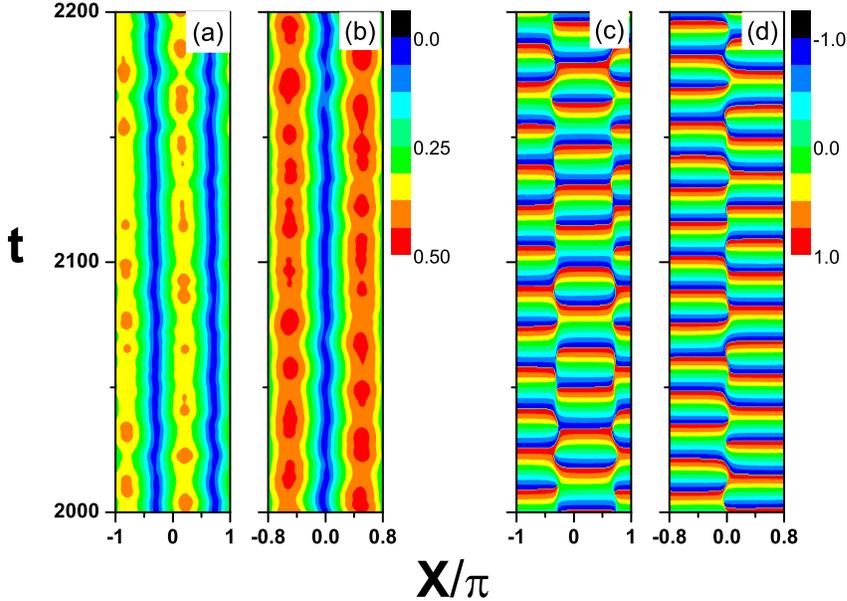}
\caption{\label{Fig3}(color online)  Phase pattern for two-cluster
chimera when the steady state is reached. Eq. (\ref{eq:12}) and
(\ref{eq:13}) is integrated using the Runge-Kutta method with
fixed time step $dt=0.005$ and the oscillator number $N=256$. (a,
c) are the modulus $R$ and the phase $\Theta/\pi$ of the complex
order parameter with $k=\pi$. (b, d) are the modulus $R$ and the
phase $\Theta/\pi$ of the complex order parameter with $k=0.8\pi$.
Other parameters: $\beta=0.1$, $A=0.995$.
 }
\end{figure}
To solve Eq. (\ref{KBFInal}), we first determinate the value of
$\Delta$. Because Eq. (\ref{KBFInal}) is left unchanged by any
rigid rotation $\Theta(x)\rightarrow\Theta(x)+\Theta_0$, we can
specify the value of $\Theta(x)$ at any point $x$ we like. We set
$\Theta(\frac{\pi}{2})=0$. Now we can get $\Delta$. Then we take
$R(x)$ and $\Theta(x)$ obtained from the dynamical simulations
 as initial guesses and use an iterative
scheme to determinate $R(x)$ and $\Theta(x)$ in function space,
behind which the idea is that the current estimates of $R(x)$ and
$\Theta(x)$ can be entered into the right-hand side of
(\ref{KBFInal}), and used to generate the new estimates appearing
on the left-hand side. Figure \ref{tukb} shows the results
obtained from Eqs.(9) and (2). To be stressed, without the
requirement of Eq.(7), the self-consistency equation for any
finite system always yields to a one-cluster chimera state, which
also evidences that the stationary two-cluster chimera state here
is not stable.

From Fig. \ref{tukb}, we could notice the order parameter of
stationary state, which has a little difference from Fig.
\ref{pi}(b) and (c) in the vicinity of the junction between two
domains. The transition of the former case performances like a
step function while the latter one is continuous. The difference
probably originates from the finite size effect in Fig. \ref{pi}.

The oscillating characteristic could be interpreted with the
assistance of the Ott-Antonsen ansatz \cite{17,18}. Following the
line in \cite{8,16,17}, we assume that there is a probability
density function $f(x,\omega,\theta,t)$ characterizing the state
of the system. This function satisfies the continuity equation
\cite{8, 16,17, 19}

\begin{eqnarray}\label{eq:5}
\frac{\partial f}{\partial
t}+\frac{\partial}{\partial\theta}(fv)=0
\end{eqnarray}

where
\begin{eqnarray}\label{eq:6}
v=\omega-\int_{-k}^{k}G(x-x^{\prime})
\int_{-\infty}^{\infty}\int_{-\pi}^{\pi}\sin(\theta-\theta^{\prime}+\alpha)f(x^{\prime},\omega,\theta^{\prime},t)d\theta^{\prime}d\omega
dx^{\prime}
\end{eqnarray}
with $\theta=\theta(x)$ and $\theta^{\prime}=\theta(x^{\prime})$.
The complex order parameter can be formulated as
\begin{eqnarray}\label{eq:7}
Z\equiv Re^{i\Theta}=\int_{-k}^{k}G(x-x^{\prime})
\int_{-\infty}^{\infty}\int_{-\pi}^{\pi}e^{i\theta^{\prime}}f(x^{\prime},\omega,\theta^{\prime},t)d\theta^{\prime}d\omega
dx^{\prime}.
\end{eqnarray}
In terms of complex order parameter $Z$, Eq. (\ref{eq:6}) can be
rewritten as
\begin{eqnarray}\label{eq:8}
v=\omega-\frac{1}{2}[Ze^{-i(\theta-\beta)}+Z^{*}e^{i(\theta-\beta)}]
\end{eqnarray}
where $Z^{*}$ denotes the complex conjugate of $Z$ and
$\beta=\frac{\pi}{2}-\alpha$. Following Ott and Antonsen \cite{17,
18}, we have
\begin{figure}
\includegraphics[width=4.0in]{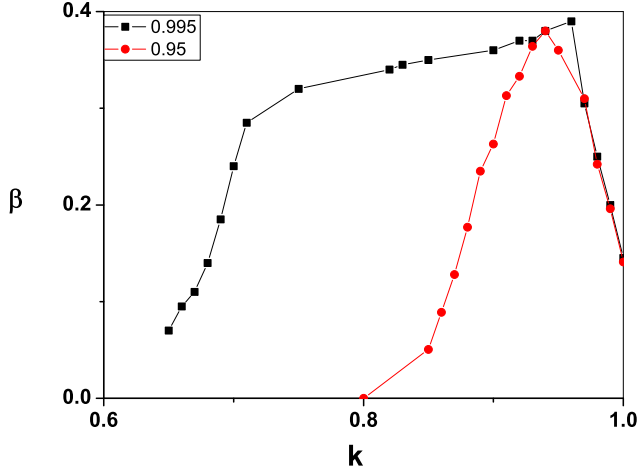}
\caption{\label{Fig4}(color online)  The boundary of the
two-cluster chimera state plotted on $k-\beta$ parameter plane.
Below the curves, the two-cluster chimera state is stable. The
square (black) symbol and circle (red) symbol are plotted at the
parameter values $A=0.995$ and $A=0.95$, respectively. }
\end{figure}

\begin{eqnarray}\label{eq:9}
f(x,\omega,\theta,t)=\frac{g(\omega)}{2\pi}\{1+\sum_{n=1}^{\infty}[(a(x,\omega,t)e^{i\theta})^n+c.c.]\}
\end{eqnarray}
where $c.c.$ is the complex conjugate of the previous term and
$g(\omega)$ is the distribution of natural frequency. In this
work, we assign all oscillators a same natural frequency
($\omega=0$), so $g(\omega)=\delta(\omega)$. Substituting
Eqs.(\ref{eq:8}) and (\ref{eq:9}) into Eqs. (\ref{eq:5}) and
(\ref{eq:7}), we obtain
\begin{eqnarray}\label{eq:10}
\frac{\partial a(x,\omega,t)}{\partial
t}=\frac{i}{2}[Z^{*}e^{-i\beta}+Ze^{i\beta}a^2]
\end{eqnarray}
\begin{eqnarray}\label{eq:11}
Z=\int_{-k}^{k}G(x-x^\prime)\int_{-\infty}^{\infty}g(\omega)a^*(x,\omega,t)d\omega
dx^{\prime}.
\end{eqnarray}
Letting $\hat{a}(x,t)=a(x,0,t)$, we have
\begin{eqnarray}\label{eq:12}
\frac{\partial \hat{a}(x,t)}{\partial
t}=\frac{i}{2}[Z^{*}e^{-i\beta}+Ze^{i\beta}\hat{a}^2]
\end{eqnarray}
\begin{eqnarray}\label{eq:13}
Z=\int_{-k}^{k}G(x-x^\prime)\hat{a}^*(x^\prime,t)dx^{\prime}.
\end{eqnarray}
\begin{figure}
\includegraphics[width=6.0in]{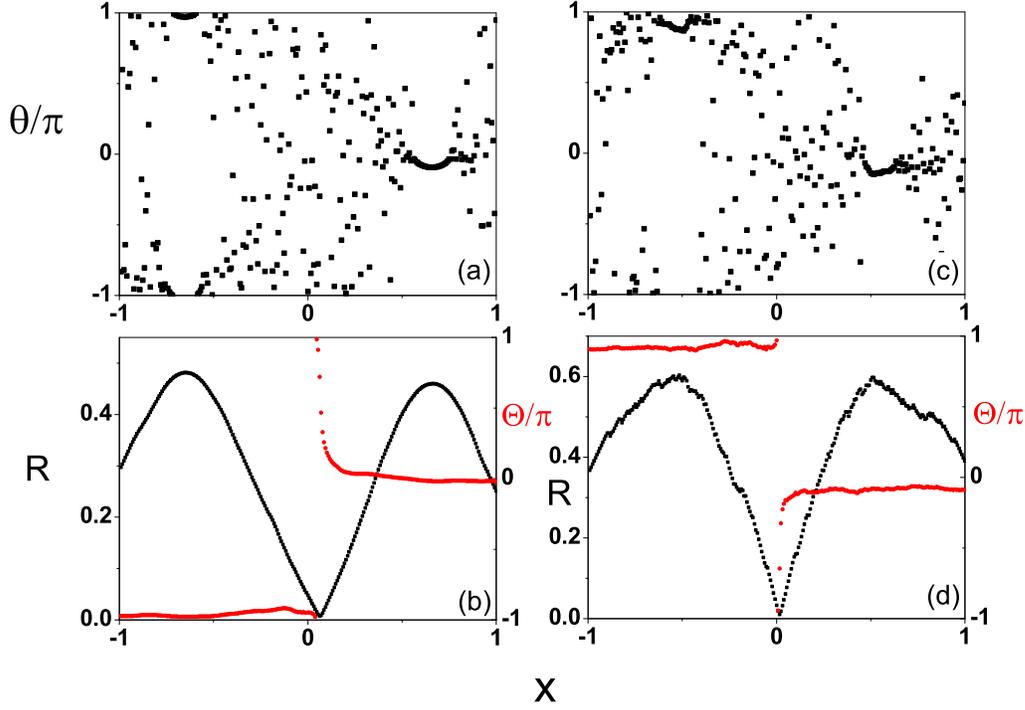}
\caption{\label{exp}(color online) Two-cluster chimera under the
condition of different types of nonlocal coupling kernel $G(x)$.
The top panels are the phase patterns for two-cluster chimera and
the bottom are the modulus $R$ (black) and the phase $\Theta$
(red) of the order parameters. (a, b) is computed under the
assumption that $G(x)$ takes exponential form with parameters:
$A=4$, $k=1$ and $\beta=0.10$ while $G(x)$ in (c, d) is steplike
function with parameters: $d=0.5$, $k=1$ and $\beta=0.10$. All
above take no-flux boundary condition and the initial conditions
are as following: $\phi(x,0)=0$ for $x<0$ and $\phi(x,0)=\pi$ for
$x>0$}
\end{figure}

By numerically simulating these two equations, we have the time
evolutions of $R(x)$ and $\Theta (x)$. The results for $k=\pi$ and
$k=0.8\pi$ are presented in Fig. \ref{Fig3}, respectively.
Clearly, the main features in Figs. (1) and (2), such as the
oscillating nature of the two-cluster chimera state, the movement
of the pattern (or the frozen pattern) of $R$ and $\Theta$ for
$k=\pi$ (or for $k=0.8\pi$), and the uniform distribution of
$\Theta (x)$ in different domains, are reproduced in Fig.
\ref{Fig3}.

Using Eqs. (\ref{eq:12}) and (\ref{eq:13}), we may probe into the
regime for the existence of the two-cluster chimera state on
$k-\beta$ parameter plane. The results are presented in Fig.
\ref{Fig4} at $A=0.995$ and $A=0.95$. As shown in this plot, the
two-cluster chimera state is not favorable at large $\beta$, and
either small $k$ or large $k$ tends to be harmful for the
two-cluster chimera state. And the smaller $A$ is, the smaller the
domain of two-clustered chimera exists. if $A<0.8$, the domain
does not exist any longer. Beyond the stable regime for the
two-cluster chimera, i.e., above the curves in Fig. \ref{Fig4},
the initial two-cluster chimera state tends to become a normal
chimera state with only one coherent cluster. The two-cluster
oscillating chimera state may be detected in coupled oscillators
with other types of non-locally coupling kernel $G(x)$ which could
be exemplified by two cases. In the first one, $G(x)$ takes an
exponentially decaying function,
$G_{exp}(x)=\frac{Ae^{-A|x|}}{2(1-e^{-Ak})}$. In the other case,
$G(x)$ takes a steplike function, $G_{step}(x,d)=\frac{1}{2d}$ for
$|x|\leq d$ and $G_{step}(x,d)=0$ for $|x|>d$. Figure 7 shows the
results for the systems with these two types of non-locally
coupling and, clearly, the two-cluster oscillating chimera states
are reproduced.

\section{Conclusion}

In summary, we study a one-dimensional system consisting of
non-locally coupled phase oscillators, which is a prototype for
studying chimera states. By numerically simulating this simplest
system, we find the existence of a two-cluster oscillating chimera
state in the absence of time delay coupling and parameter
heterogeneity. The numerical results are confirmed by the
theoretical analysis based on the self-consistency treatment and
the Ott-Antonsen ansatz.

\section*{Acknowledgments}

The work was supported by National Natural Science Foundation of
China under Grant No. 90921015 and No. 10775022.

\end{document}